\documentclass[aps,prl,twocolumn,showpacs]{revtex4}

\usepackage{graphicx}
\usepackage{amsmath,amscd,amssymb}
\newcommand \lan {\langle}
\newcommand \ran {\rangle}
\newcommand \kt {\tilde \kappa}

\begin{document}

\title{Thermal Denaturation of Fluctuating DNA Driven by Bending Entropy}
\author{J. Palmeri, M. Manghi and N. Destainville}

\affiliation{Laboratoire de Physique Th\'eorique, Universit\'e de Toulouse, CNRS, 31062 Toulouse, France}

\pacs{87.10.+e, 87.15.Ya, 82.39.Pj}

\date{\today}

\begin{abstract}
A statistical model of homopolymer DNA, coupling internal base pair
states (unbroken or broken) and external thermal chain fluctuations,
is exactly solved using transfer kernel techniques. The dependence
on temperature and DNA length of the fraction of denaturation
bubbles and their correlation length is deduced. The {\em thermal
denaturation transition} emerges naturally when the chain fluctuations are integrated out
and is driven by the
difference in bending (entropy dominated) free energy between broken
and unbroken segments. Conformational properties of DNA, such as
persistence length and mean-square-radius, are also explicitly
calculated, leading, e.g., to a coherent explanation for the
experimentally observed {\em thermal viscosity transition}.
\end{abstract}

\maketitle

Double-stranded DNA (dsDNA) is made up of two intertwined
interacting semi-flexible single-strand DNA (ssDNA) chains. Over fifty years ago it was
recognized that the intracellular unwinding
of DNA at physiological temperature has as counterpart the thermally
induced denaturation above physiological temperature of purified DNA
solutions where dsDNA completely separates into two ssDNA. Despite
the differences between the two mechanisms, this observation has led
to an intensive study of thermal
denaturation~\cite{poland,wartmont}. The stability of dsDNA at
physiological temperature is due to the self-assembly of neighboring
base pairs within a same strand {\em via} base-stacking interactions
and of both strands {\em via} hydrogen bonds between complementary
bases. The bonding energy is, however, on the order of $k_BT$ (thermal energy)~\cite{pincet} and thermal fluctuations can
lead, even at physiological temperature, to local and transitory
unzipping of dsDNA~\cite{wartmont}. The cooperative opening of
consecutive base pairs leads to denaturation bubbles and the melting
temperature, $T_m$, above which bubbles proliferate, depends on
sequence, chain length, and ionic strength.
Experiments show, for example, that there exist a bubble initiation barrier of $\sim 10 k_BT$ and free energy cost of $\sim 0.1 k_BT$ for breaking an additional base pair in an existing A-T bubble~\cite{krueger}.
A detailed understanding of equilibrium~\cite{wartmont} and dynamical~\cite{metzler}
properties of DNA in solution is still being sought and a consensus
concerning the physical mechanism behind the denaturation
transition has not yet been reached.

A variety of mesoscopic models have been proposed to account for the
thermodynamical properties of denaturation bubbles in DNA. They
range from i) simple effective Ising-like two-state
models~\cite{wartmont} to more detailed ones such as ii) loop
entropy models (with or without chain
self-avoidance)~\cite{poland,wartmont,peliti,computer}, and iii)
non-linear phonon models, where the shape of the interaction
potential between base pairs is more precisely taken into
account~\cite{Peyrard98,jeon}. To get a transition in models i) and
ii), an effective temperature dependent base-pair chemical potential
must be inserted by hand. For finite chains, type (ii) models simply
refine the sharpness of the transition~\cite{wartmont}, but do not
attempt to provide a deeper explanation of the physical mechanism --
our aim here. For type (iii) models it has been shown that there can
be a denaturation transition analogous to interface unbinding, due to a
gain in translational entropy. If,
however, physically reasonable values for the model parameters are
used~\cite{gao,Peyrard98,jeon}, $T_m$ appears to be much too high
and the transition width much too large.

It has been shown experimentally that dsDNA is two orders of
magnitude stiffer than ssDNA at normal salt concentration. We show
in this Letter that taking into account this difference in
\textit{bending rigidity} provides a novel physical explanation for
the denaturation mechanism and leads to realistic values for
transition temperatures and widths. Despite important recent
advances in understanding the crucial role played by DNA
\textit{bending rigidity} in explaining
force-extension~\cite{Marko04,Nelson03} and cyclization~\cite{menon}
experiments, its importance has not yet been clearly elucidated in
the context of denaturation. We show, \textit{via} a
well-defined coupled Ising-Heisenberg statistical model, that an
entropy driven denaturation transition emerges by integrating out
chain fluctuations, due to the entropic lowering of the
energetic barrier for bubble nucleation. This minimal model neglects
all other (residual) interactions between bases arising from,
\textit{e.g.}, electrostatics~\cite{korolev},
self-avoidance~\cite{peliti}, and helical twist~\cite{benham}.

We begin by considering a worm like chain (WLC) Hamiltonian $H[{\bf
r}_1, {\bf r}_2]$ for two interacting ssDNA homopolymers of length $L$:
\begin{eqnarray}
H & = &  \frac12 \sum_{i = 1}^{2} \int_0^L d s \left[ \frac32
\kappa_b(\boldsymbol{\rho}(s)) \ddot{\bf r}_i^2(s) +
\frac32\kappa_s(\boldsymbol{\rho}(s)) \dot{\bf r}_i^2(s)\right] \label{Ham}  \nonumber\\
  & &  + \int_0^L d s\, V(\boldsymbol{\rho}(s))), \label{Hini}
\end{eqnarray}
where $s$ is the curvilinear index, ${\bf
r}_i(s)$ is the position of chain $i$ at base position $s$,
and  $\boldsymbol{\rho}(s) \equiv {\bf r}_1 (s)- {\bf r}_2 (s)$ is
the relative (internal) coordinate ($\beta = 1/k_BT$ and $\dot{\bf
r}(s)=\frac{\partial {\bf r}}{\partial s}$).  The coefficient $\kappa_b$ is
a {\em bending} elastic modulus that is proportional to the short
distance cut-off $\ell_0 \approx 0.34$~nm (the
monomer length) and $\kappa_s(\boldsymbol{\rho})=1/[\beta^2\kappa_b(\boldsymbol{\rho})]$. In order to account for the enhanced stiffness of
dsDNA, $\kappa_b$ must depend on $\boldsymbol{\rho}(s)$, e.g., $\kappa_b(\boldsymbol{\rho}) =
\kappa_{b} ^{ss} + \Delta\kappa_{b} e^{-(\rho - \rho_0)/\lambda}$, where $\Delta\kappa_{b} \equiv \kappa_{b}^{ds}/2 -
\kappa_{b}^{ss}>0$ and $\lambda \sim 1\AA$ is the range~\cite{gao}. The persistence length of ssDNA is $\ell_p^{ss} =\beta
\kappa_{b}^{ss}\approx1$~nm  and that of dsDNA is $\ell_p^{ds} =\beta
\kappa_{b}^{ds}\approx50$~nm (at 300~K and physiological ionic strength). The
potential $V(\rho)$ accounts for both the effective short range
hydrogen bonding interaction between complementary bases at the same
$s$ and part of the stacking interaction between neighboring bases;
a convenient form is the Morse potential: $V({\bf \rho}) = (D/\ell_0)v((\rho-\rho_0)/\lambda)$ where
$v(x) = e^{ - 2x}  - 2e^{ - x} $, leading to a well depth of
$D/\ell_0$ at $\rho_0$~\cite{Peyrard98}.

At high temperature, $\rho(s)-\rho_0 \gg \lambda$, and
therefore the system decouples into two semi-flexible
non-interacting ssDNA chains. After introducing the center-of-mass
(external) coordinate, ${\bf X}(s) \equiv [{\bf r}_1 (s)+ {\bf r}_2
(s)]/2$, the partition function can be rewritten as a sum over
center-of-mass and relative configurations: $\mathcal{Q} =
\int\mathcal{D}\boldsymbol{\rho}\mathcal{D} {\bf X} \exp
\{-\beta(H_{\rm ext}[{\bf X}, \boldsymbol{\rho}] + H_{\rm
int}[\boldsymbol{\rho}])\}$, where~\cite{footnote}
\begin{eqnarray}
H_{\rm int} &=&  \frac38  \int_0^L d s \left[
\kappa_b(\boldsymbol{\rho})  \ddot{\boldsymbol{\rho}}^2 +
\kappa_s(\boldsymbol{\rho})   \dot{\boldsymbol{\rho}}^2 +
V(\boldsymbol{\rho})\right] \label{Hint}\\
H_{\rm ext} &=& \frac32 \int_0^L d s \left[
\kappa_b(\boldsymbol{\rho}) \ddot{\bf X}^2 +
\kappa_s(\boldsymbol{\rho}) \dot{\bf X}^2 \right]. \label{Hext}
\end{eqnarray}

We integrate over ${\bf X}$ to obtain an effective model for
$\boldsymbol{\rho}$, $ \mathcal{Q} =
\int\mathcal{D}\boldsymbol{\rho} \exp \{ - \beta H_{\rm
eff}[\boldsymbol{\rho}] \}$, where $ H_{\rm eff}[\boldsymbol{\rho}]
= H_{\rm int}[\boldsymbol{\rho}] + \mathcal{F}_{\rm
ext}[\boldsymbol{\rho}]$ with $\mathcal{F}_{\rm
ext}[\boldsymbol{\rho}] = -k_B T \ln \left[ \int\mathcal{D} {\bf X}
\exp \left\{ - \beta H_{\rm ext}[{\bf X}, \boldsymbol{\rho}]
\right\} \right]$. The {\em external} free energy $\mathcal{F}_{\rm
ext}[\boldsymbol{\rho}]$ at frozen $\boldsymbol{\rho}$ can be
evaluated by introducing the local center-of-mass tangent vectors
${\bf t}=\dot{\bf X}$ and then changing variables to $\tilde{\bf t}=
{\bf t}/\sqrt{\kappa_b(\boldsymbol{\rho})}$. There are two
contributions to $\mathcal{F}_{\rm ext}$: one will renormalize the
second term in $H_{\rm int}$ and the other will renormalize the
potential to $V_{R}$. The latter mainly gives rise to a purely
entropic barrier favoring bubbles,
\begin{equation}\label{vr}
V_{R}(\boldsymbol{\rho}) =  V(\boldsymbol{\rho}) +
 \frac32 \frac{k_BT}{\ell_0} \ln
\left[ \frac{\kappa_b(\boldsymbol{\rho})}{\kappa_b^{ss}}\right],
\end{equation}
which lowers the well depth from $D$ to $D_{R} \approx   D -
(3/2) k_{B}T\ln \left[ \kappa_{b}^{ds}/(2 \kappa_{b}^{ss}) \right]$.
The entropic bending contribution can be extremely important when
$D$ is in the commonly accepted range of $1<D/k_BT<5$ at $T =
350~\mathrm{K}$ and $\kappa_{b}^{ds}/(2 \kappa_{b}^{ss}) \simeq 25$.
Scaling arguments then show that the melting temperature  gets
reduced by a factor $\sim 2$ down to experimental
values~\cite{Peyrard98,longpaper}.

In order to illustrate the above mechanism in more detail, we
introduce a discretized, exactly soluble, version of the above
model, which captures the essential physics. We map the external
tangent vector, ${\bf t}(s)$, to ${\bf t}_i$ ($s=\ell_0 i$) and an
internal variable $1-2\, \Theta(\rho(s)-\rho_0-\lambda)$ (with
$\Theta$ the step function) to an Ising variable: $\sigma _i =1$ for
an unbroken bond (state A) and $\sigma _i =-1$ for a broken one~(B).
Each link vector can be denoted by the solid angle $\Omega _i =
(\theta _i ,\phi _i )$. The energy $H[\sigma _i ,{\bf{t}}_i ]$
 of a state is
\begin{eqnarray}
H & = & \sum_{i = 1}^{N - 1}  \, {\tilde \kappa}_{i,i + 1} (1 -{\bf
t}_i \cdot {\bf t}_{i+1})+ H_{\rm I}(\tilde J,\tilde K,\tilde \mu), \label{H}\\
H_{\rm I} &=& -\sum_{i = 1}^{N-1} \, \left[\tilde J\sigma_i
\sigma_{i+1}+\frac{\tilde K}2 (\sigma_i + \sigma_{i+1})\right]  - \tilde \mu \sum_{i =1}^N \,\sigma_i. \nonumber
\end{eqnarray}
Thus $\beta H$
contains only the dimensionless parameters $\kappa _{i,i+1}  \equiv \beta
{\tilde \kappa}_{i,i+1} $, $J \equiv \beta \tilde J$,  $\mu\equiv
\beta \tilde \mu$, and $K\equiv \beta \tilde K$.
This type of model was introduced in the context of helix-coil
transitions in 2D~\cite{john} and later used in various forms to
study DNA force-extension and
cyclization~\cite{Nelson03,Marko04,menon}. The first term in
$H$, corresponding to $H_{\rm ext}$ in the continuum model, is the
bending energy of a discrete WLC with a local rigidity
$\kappa _{i,i+1}=\kappa_A=\beta\kappa_b^{ds}/\ell_0$ for a nearest
neighbor link of type A-A, $\kappa_B=2\beta\kappa_b^{ss}/\ell_0$ for
B-B and $\kappa_{AB}$ for A-B. In the Ising part, $H_\mathrm{I}$, the first term  mimics the gradient
terms in Eq.~(\ref{Hint}) and accounts for the local destacking
energy~\cite{Peyrard98}. The second term accounts for the difference in
stacking energy between a segment of dsDNA and a
denaturation bubble. The third term corresponds to the energy
($2\tilde\mu$) required to break a base pair, contributing to $D$ in
the continuum model. There is evidence that $\tilde K\ll\tilde \mu$, which  justifies the choice of $\tilde K = 0$ adopted below~\cite{goddard}. The bare parameters of the internal (Ising) system
$\tilde J$, $\tilde \mu$, and $\tilde K$ are taken to be independent
of temperature and bubble loop entropy is not explicitly included;
the melting transition is therefore driven only by the difference in
bending rigidity.  The partition function $ Z =  \sum_{\{ \sigma
_i\} }\int\prod_i \frac{d\Omega _i }{4\pi} e^{- \beta H[\sigma
_i,{\bf t}_i ]}$ in transfer kernel form is
\begin{eqnarray}
Z &=&  \sum_{\{ \sigma _i = \pm 1\} }  \, \prod_{i = 1}^N  \,\int
\frac{d\Omega _i }{4\pi} \langle {V|\sigma _1 } \rangle \langle
\sigma _1 |\hat P(\Omega _1 ,\Omega _2 )|\sigma _2\rangle
\cdots \nonumber\\
& &\cdots \langle \sigma _{N - 1} |\hat P(\Omega _{N - 1} ,\Omega _N
)|\sigma _N \rangle \langle {\sigma _N |V}\rangle,\label{partfunc}
\end{eqnarray}
where $| V \rangle  = ( {e^{\mu /2} }  ,  {e^{ - \mu /2} }) $ is the
end vector, which enters to account for the free chain boundary
conditions, and $\hat{P} (\Omega _{i,} ,\Omega _{i+1} )$ is the
transfer kernel given by
\begin{equation}
\hat{P} =\left(
\begin{array}{*{20}c}
   {e^{\kappa_A [\cos\gamma_i-1] + J + \mu + K} } & {e^{\kappa_{AB}
[\cos\gamma_i - 1] - J} }  \\
   {e^{\kappa_{AB} [\cos\gamma_i-1] - J} } & {e^{\kappa_B [\cos\gamma_i - 1] + J - \mu - K } }  \\
\end{array} \right)
\end{equation}
with $\cos\gamma_i={\bf t}_i\cdot{\bf t}_{i+1}$. The A and B
states form the canonical base, $|A \rangle = | +1\ran= (1,0)$ and
$| B \rangle = | -1\ran = (0,1)$. Thanks to the rotational symmetry
of the bending energy we can again integrate out the chain, leading
to an effective Ising model with a free energy, $H_{{\rm I},0}$,
containing renormalized parameters: $ Z = e^{-(N-1)\Gamma _0}
\sum_{\{ \sigma _i \} } \,e^{ -\beta H_{{\rm I},0} [\sigma _i ]} $ where $H_{{\rm I},0}\equiv H_{\rm I}(\tilde J_0,\tilde K_0,\tilde \mu)$
with $J_0 \equiv J
-[G_0(\kappa_A)+G_0(\kappa_B)-2G_0(\kappa_{AB})]/4$ and $K_0 \equiv
K-[G_0(\kappa_A)-G_0(\kappa_B)]/2 \equiv K - \Delta G_0^{AB}/2$
the renormalized Ising parameters, and
$\Gamma_0\equiv[G_0(\kappa_A)+G_0(\kappa_B)+2G_0(\kappa_{AB})]/4$.
These parameters depend on chain rigidities through $G_0 (\kappa )$
which is the  free energy of a single joint (two monomer) subsystem
with rigidity $\kappa$: $ G_0(\kappa)= -\ln\{\int
\frac{\mathrm{d}\Omega}{4\pi}\exp[\kappa (\cos(\theta )-1)]\} =
\kappa  - \ln[\sinh(\kappa )/\kappa]$. The renormalized quantity $2
L_0\equiv 2(\mu + K_0)$ corresponds to $\beta D_{\rm R}$ in the continuum model. If the
bending free energy gain in opening one link, $\Delta G_0^{AB} $, is
greater than the intrinsic energy cost, $ 2(\mu+K)$, of opening an
interior bond, then $L _0$ becomes negative, signaling a change in
stability of the A and B states. In the limit of high $\kappa_A$ and $\kappa_B$, the entropic
contribution dominates: $G_0(\kappa) \approx- S_0(\kappa) /k_{B}$
$\approx \ln(2\kappa )$. The discrete model then reduces to the
continuum one and $\Delta G_0^{AB} \approx \ln
(\kappa_{A}/\kappa_{B})$ corresponds to the correction appearing in
$\beta D_{\rm R}$. The difference between $L_0$ and $\mu$ when $K_0
\neq 0$  creates an end-interior asymmetry that plays an important
role in finite size effects.

The Ising partition and correlation functions are obtained using
transfer matrix techniques. The eigenvectors, $|0,\pm\ran$, and the
eigenvalues, $\lambda_{0,\pm}= e^{J_0-\Gamma_0}[\cosh(L_0) \pm
(\sinh^2(L_0) + e^{-4J_0})^{1/2}]$, of the effective Ising
transfer matrix allow us to calculate the (dimensionless) free
energy per Ising spin of the coupled system, $F=-\ln Z/N$, where $Z
=  \sum_{\tau=\pm} \langle V | 0,\tau \rangle^2 \lambda
_{0,\tau}^{N-1}$. The average of the internal state variable is
$\langle c\rangle = \langle \sum_{i = 1}^N \,\sigma _i /N\rangle=-
\partial F/\partial \mu$, from which the fractions of A and B links,
$\varphi _A =(1 + \langle c \rangle )/2$ and $\varphi
_B=1-\varphi_A$, can be derived. The melting temperature $T_m$ is
then defined by $\varphi_B(T_m)=1/2$. When $N \to \infty$, $\langle c
\rangle$ gets simplified to
\begin{equation}
\left\langle c \right\rangle_\infty \equiv \frac{\partial \ln\lambda
_{0,+ }}{\partial \mu} = \frac{\sinh(L_0)}{\left[\sinh^2(L_0) +
e^{-4J_0}\right]^{1/2}}. \label{c}
\end{equation}
If $L_0$ vanishes at a temperature, $T_m^{\infty}$, sufficiently low
for the cooperativity, or loop initiation, factor, $\sigma \equiv
e^{-4J_0}$, in the denominator to be small, then the system will
undergo a melting transition: $\langle c \rangle_\infty$ will
sharply cross-over from $+1$ for $T < T_m^{\infty}$ (pure A state)
to $-1$ for $T > T_m^{\infty}$ (pure B state). \textit{Contrary to
previous Ising-type models, the melting temperature is not put in by
hand but emerges naturally from} $L_0=0$. In Eq.~(\ref{c})
$e^{-4J_0}$ determines the width of the transition region: $\Delta
T_m^{\infty} \equiv 2 |\partial \langle c\rangle_\infty/\partial T
|_{T_m^\infty}^{-1} \approx (2\,k_B (T_m^{\infty})^2/\tilde\mu) \exp
[-2\,J_0(T_m^{\infty})]$. In the limit $N,i \to \infty$, the
influence of end-monomers disappears and  the Ising correlation
function reduces to $\lan (\sigma _i-\langle c\rangle_\infty)(\sigma
_{i + r} - \langle c\rangle_\infty)\ran \to (1-\langle
c\rangle_\infty^2) \exp(-r/\xi_{\rm I})$, where $\xi_{\rm
I}=-\ln^{-1}(\lambda_{0,-}/\lambda_{0,+})$ is the Ising correlation
length, the typical size of B (A) domains below (above) $T_m$.

The tangent-tangent correlation function, $ \langle {\bf{t}}_i \cdot
{\bf{t}}_{i + r} \rangle$, is obtained using the full transfer
kernel method~\cite{longpaper}, which requires solving a spinor
eigenvalue problem: $\hat P|\hat \Psi \rangle  = \lambda |\hat \Psi
\rangle$, where $|\hat
\Psi(\Omega)\rangle=(\Psi_{+1}(\Omega),\Psi_{-1}(\Omega) )$. We find
eigenvalues, $\lambda_{l,\tau}$, labeled by $l=0,\ldots,\infty$ and
$\tau=\pm$ with the same form as for $l=0$ given above, but now
where $G_0$ in the renormalized parameters is replaced by
$G_l(\kappa)=\kappa-\ln[\kappa^l
(\mathrm{d}/\kappa\mathrm{d}\kappa)^l
(\sinh(\kappa)/\kappa)]$~\cite{joyce}. The eigenspinors are $|\hat
\Psi_{l,m;\tau } (\Omega) \rangle  = \sqrt{4\pi}Y_{lm}(\Omega)|
l,\tau\rangle$ with $Y_{lm}(\Omega)$ the spherical harmonics. The
transfer kernel can be expanded in terms of the eigenspinors $\hat P
= \sum_{l,m,\tau}\lambda_{l,\tau} |\hat \Psi_{l,m;\tau } \rangle
\langle \hat \Psi_{l,m;\tau }|$ and then be used to calculate the
correlation functions. In the  limit $N,i \to \infty $, the
expression for the tangent-tangent correlation function simplifies
to
\begin{equation}
\langle {\bf{t}}_i  \cdot {\bf{t}}_{i + r}\rangle \underset{N,i
\to\infty}{\to} \sum_{\tau=\pm} \langle 1,\tau | 0,+ \rangle^2 \exp
[- r/\xi _{1,\tau}^p] \label{tangentcorrNiinf}
\end{equation}
which reveals the importance of the two persistence lengths,
$\xi_{1,\pm}^p=-\ln^{-1}(\lambda _{1, \pm } /\lambda _{0, + })$
(units of $\ell_0$). It is not possible to extract from $\langle
{\bf{t}}_i \cdot {\bf{t}}_{i + r}\rangle$ one unique length for the
whole range of $T$, because the weights associated with
$\xi_{1,\pm}^p$ strongly vary with $T$.

\begin{figure}[t]
\includegraphics[height=9cm]{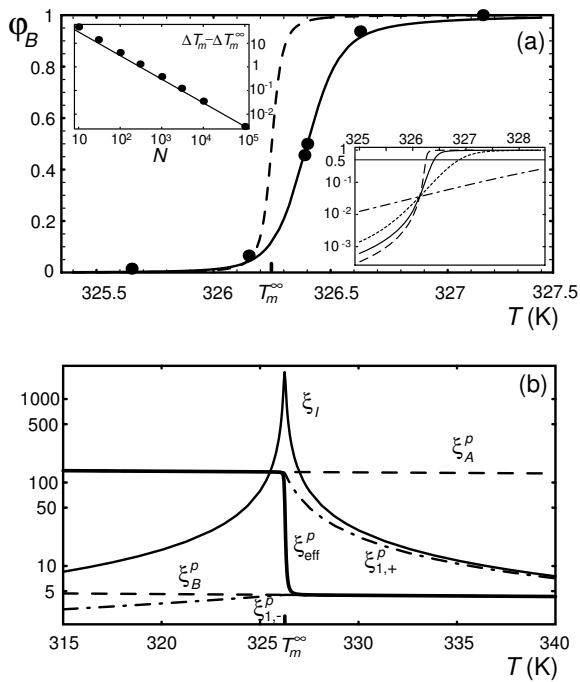}
\caption{(a)~ Melting curves (Fraction of broken base-pairs vs.
temperature) for poly\textit{d}A-poly\textit{d}T (data for 0.1 SSC ($= 0.015M$ NaCl $+ 0.0015M$ sodium citrate, pH 7.0),  
$N=$1815~\cite{wartmont}). The solid line represents the theoretical
result for $\tilde\mu=1.64\,k_BT_m$, $\tilde{J}=3.35\,k_BT_m$
($T_m=326.4$~K). The broken line corresponds to $N\to\infty$ (same
parameter values). Lower inset: melting curves for $N= 100, 500,
1815, \infty$ (in decreasing order at low temperture, $T < 326$~K);
Upper inset: Model results for the shift in transition width $\Delta
T_m-\Delta T_m^{\infty}$ vs. polymer length~\cite{footnote2}. (b) Temperature
variation ($N\to\infty$) of the Ising correlation length, $\xi_{\rm
I}$; persistence lengths of the coupled system, $\xi^p_{\rm eff}$
and $\xi _{1,\pm}^p$; and of the pure chains, $\xi^p_{A,B}$ (in
units of $\ell_0$). At $T_m^{\infty}$, the effective persistence
length, $\xi^p_{\rm eff}$, rapidly crosses over from $\xi^p_{A}$ to $\xi^p_{B}$.}
\label{fig1}
\end{figure}

We now compare the discrete model predictions with
experiment~\cite{wartmont} by focusing on the melting profile,
$\varphi_B(T)$, of a  synthetic nonalternating homopolynucleotide,
poly\textit{d}A-poly\textit{d}T.  A typical profile for a
homopolynucleotide has a sigmoidal shape characterized by $T_m$ and
$\Delta T_m$.  Of the five independent parameters that appear in the
theory when $K = 0$, three are determined experimentally
(polymerization index $N$ and bending moduli assuming
$\kappa_{AB}=\kappa_A$~\cite{longpaper}) and two ($\tilde{\mu}$ and
$\tilde{J}$) are determined by fitting the model to experiment,
hence $T_m = T_m(\tilde{\mu}, \tilde{J};N,\kappa_A,\kappa_B)$.
Figure~\ref{fig1}a shows $\varphi_B(T)$ for DNA of molecular weight
1180~kDa~\cite{wartmont}. From the known persistence lengths we
obtain $\kappa_A=147$ and $\kappa_B=5.54$ at 300~K. The solid line
corresponds to our model fit with $\tilde{\mu}=4.46$~kJ/mol and
$\tilde{J}=9.13$~kJ/mol, leading to $T_m=326.4$~K. The fitted value
for $2\tilde{\mu}$ is close to the experimental energy of 10.5
kJ/mol needed to break an A-T link~\cite{pincet}. The value for
$\tilde{J}\sim 2 \tilde{\mu}$ is also consistent with the idea that
destacking energy makes the dominant contribution to DNA
stability~\cite{wartmont}. The renormalized cooperativity parameter
at $T_m$ is $\tilde{J}_0=11.5\,{\rm kJ/mol}>\tilde J$. The model fit
thus leads to parameter values in accord with experiment (in
reality, the fitted values of $\tilde{\mu}$ and $\tilde{J}$
implicitly compensate for effects like loop entropy explicitly left
out of the model~\cite{wartmont,longpaper}). In Fig.~\ref{fig1}a,
the curve corresponding to $N\to \infty$ is shown for the same
parameter values. In this case, $\varphi_B(T)$ is given by
Eq.~(\ref{c}) and $T_m^{\infty}$ is obtained from $L_0=0$: $k_B
T_m^{\infty} \simeq 2(\tilde\mu+\tilde K)/\ln (\kt_A/\kt_B)$. In
this limit the transition width is non-zero, due to the finite
cooperativity parameter: $\Delta T_m^{\infty} \propto  \xi_{\rm
I}^{-1}(T_m^{\infty}) = 2\exp[-2J_0 (T_m^{\infty})]$. Since $\xi
_{\rm I}(T_m^{\infty}) \sim 2000$, typical helix and bubble domains
are flexible within a small window about $T_m^{\infty}$. When $N$
decreases, the width increases (Fig.~\ref{fig1}a) roughly as $\Delta
T_m(N)-\Delta T_m^{\infty} \sim N^{-1}$. Hence even for a long
polymer ($N\sim 10^3$), finite size effects are non-negligible, in
agreement with experiments~\cite{libch}. Then the nature of end
monomers becomes important, as confirmed in Fig.~\ref{fig1}a, which
shows how short chains begin to melt by end-unwinding at lower
temperatures (the trend that $T_m$ increases with decreasing $N$
will likely be reversed when loop entropy is
included~\cite{poland,longpaper}). Our model predictions for
experimentally accessible A-T pair quantities  are in agreement with
accepted values~\cite{metzler,krueger}: i)~$\sigma \approx 10^{-7}$
at $T_m$;  and at physiological temperature ii)~ an interior single
base-pair opening probability of  $10^{-6}$ with a bubble initiation
barrier of $17 k_B T$ , and iii) a free energy of $0.18 k_B T$ for
breaking an additional base-pair in an already existing bubble.

In contrast to purely Ising-type models, included in the
predictions of our theory are mechanical and structural features of
the fluctuating chain, such as persistence length or
mean-square-radius, $R \equiv \langle {\bf R}^2\rangle^{1/2}$ with
${\bf R}=\ell_0\sum_i {\bf t}_i$.  From the expression for $R$ in
the limit $N \to \infty$, we can define an effective persistence
length, $\xi^p_{\rm eff}$: $R^2 \simeq (2 \ell_0^2 N) \xi^p_{\rm eff} = (2 \ell_0^2 N)
\sum_{\tau=\pm}\langle 0, + | 1,\tau \rangle^2 \xi _{1,\tau }^p$.
Due to the coupling between bending and internal states, the
respective weights $\lan 0,+|1,\pm\ran^2$ associated with each
correlation length change abruptly at $T_m$ (cf. Fig.~\ref{fig1}b). Since the
transition is very abrupt, it should also be possible to observe it
experimentally  by measuring the radius of gyration by tethered
particle motion~\cite{pouget}, light scattering, or viscosity
experiments. For instance, since the \textit{relative viscosity} is
proportional to $R^3$, it should clearly exhibit an abrupt thermal
transition. Such a transition has indeed been observed for the
viscosity of synthetic homopolynucleotide solutions~\cite{inman}, in qualitative
agreement with Fig.~\ref{fig1}b.
Incorporating bending rigidity into DNA denaturation models  thus
allows us to make explicit predictions for both melting profiles and
DNA mean-size dependent quantities. It will be of great interest
 to both probe such effects  by carrying out  experiments on DNA
homopolymers and other biopolymers undergoing helix-coil
transitions and extend our theory to heteropolymers, mechanical denaturation, and DNA dynamics.


\begin{thebibliography}{99}
\bibitem{footnote}
In reality, $\kappa_s$ in Eq. (\ref{Hint}) should include an enthalpic shear
modulus term accounting for DNA stacking energy~\cite{Peyrard98}.
\end{thebibliography}
\end{document}